\documentclass[journal=jacsat,manuscript=article]{achemso}

\usepackage{chemformula} 
\usepackage[T1]{fontenc} 

\usepackage{etoolbox}

\makeatletter
\renewcommand*{\acs@author@fnsymbol@symbol}[1]{
    \ifcase #1 *\or
    1\or
    2\or
    3\or
    4\or
    5\or
    6\or
    7\or
    8\or
    9\or
    10
    \fi
}
        
\renewcommand*\acs@contact@details{
    {\sffamily *\,E-mail: \acs@email@list }%
    \acs@number@list
}           

\patchcmd{\acs@address@list@auxii}
{\acs@author@fnsymbol{\acs@affil@marker@cnt}}
{\textsuperscript{\acs@author@fnsymbol{\acs@affil@marker@cnt}}}
{}{}

\patchcmd{\acs@address@list@auxii}
{{\acs@author@fnsymbol{\acs@affil@marker@cnt}\@nameuse{@altaffil@\@roman\@tempcnta}\par}}
{{\textsuperscript{\acs@author@fnsymbol{\acs@affil@marker@cnt}}\@nameuse{@altaffil@\@roman\@tempcnta}\par}}
{}{}        

\makeatother    
\author{Wantong Huang}
\affiliation{State Key Laboratory of Low-Dimensional Quantum Physics, Department of Physics, Tsinghua University, Beijing 100084, China}

\author{Haicheng Lin}
\affiliation{State Key Laboratory of Low-Dimensional Quantum Physics, Department of Physics, Tsinghua University, Beijing 100084, China}

\author{Yuguo Yin}
\affiliation{State Key Laboratory of Low-Dimensional Quantum Physics, Department of Physics, Tsinghua University, Beijing 100084, China}

\author{Cheng Zheng}
\affiliation{State Key Laboratory of Low-Dimensional Quantum Physics, Department of Physics, Tsinghua University, Beijing 100084, China}

\author{Wei Chen}
\affiliation{State Key Laboratory of Low-Dimensional Quantum Physics, Department of Physics, Tsinghua University, Beijing 100084, China}

\author{Lichen Ji}
\affiliation{State Key Laboratory of Low-Dimensional Quantum Physics, Department of Physics, Tsinghua University, Beijing 100084, China}

\author{Jack Hughes} 

\affiliation{College of Engineering and Physical Sciences, Khalifa University, PO Box 127788, Abu Dhabi, UAE}

\author{Fedor Kusmartsev} 

\affiliation{College of Engineering and Physical Sciences, Khalifa University, PO Box 127788, Abu Dhabi, UAE}
\alsoaffiliation{Physics Department, Loughborough University, LE11 3TU, UK}

\author{Anna Kusmartseva}
\affiliation{Physics Department, Loughborough University, LE11 3TU, UK}

\author{Qi-Kun Xue}
\affiliation{State Key Laboratory of Low-Dimensional Quantum Physics, Department of Physics, Tsinghua University, Beijing 100084, China}
\alsoaffiliation{Frontier Science Center for  Quantum Information, Beijing, 100084, China}

\author{Xi Chen}
\email{xc@mail.tsinghua.edu.cn}
\affiliation{State Key Laboratory of Low-Dimensional Quantum Physics, Department of Physics, Tsinghua University, Beijing 100084, China}
\alsoaffiliation{Frontier Science Center for  Quantum Information, Beijing, 100084, China}

\author{Shuai-Hua Ji}
\email{shji@mail.tsinghua.edu.cn}
\affiliation{State Key Laboratory of Low-Dimensional Quantum Physics, Department of Physics, Tsinghua University, Beijing 100084, China}
\alsoaffiliation{Frontier Science Center for  Quantum Information, Beijing, 100084, China}
\alsoaffiliation{Collaborative Innovation Center of Quantum Matter, Beijing 100084, China}

\title{Landau-Level Quantization and Band Splitting of FeSe Monolayers Revealed by Scanning Tunneling Spectroscopy}

\keywords{\textbf{Keywords:} Rashba-type spin-orbit coupling, Landau level sepctroscopy, band splitting, nonparabolic electron band}

\begin{document}








\begin{abstract}
Two-dimensional (2D) superconductors that reside on substrates must be influenced by Rashba spin-orbit coupling (SOC). The intriguing effect of Rashba-type SOCs on iron-based superconductors (IBSs) has remained largely a mystery. In this work, we unveil modified Landau-level spectroscopy and the intricate band splitting of FeSe monolayers through the precision of scanning tunneling spectroscopy, which unequivocally demonstrates the presence of Rashba SOC. The discovery sheds light on a nonparabolic electron band at the X/Y point, displaying a distinctive Landau quantization behavior characterized by $E_n\propto(nB)^{4/3}$. The theoretical model aligns with our experimental insights, positing that the k$^4$-term of the electron band becomes predominant and profoundly reshapes the band structure. Our research underscores the pivotal role of the Rashba SOC effect on 2D superconductors and sets the stage to probe new quantum states in systems with remarkably low carrier concentrations.
\end{abstract}
{\textbf{Keywords:} Rashba-type spin-orbit coupling, Landau level sepctroscopy, band splitting, nonparabolic electron band}

The out-of-plane inversion symmetry of thin films is broken by the substrates, inducing the Rashba-type spin-orbit coupling (SOC). Consequently, the spin degeneracy of the electronic bands is lifted. In superconductors, the spin splitting can lead to exotic phenomena, such as  mixed singlet-triplet superconducting pairing  \cite{PhysRevB.80.024504,PhysRevLett.87.037004,PhysRevB.67.020505}, enhanced critical magnetic field \cite{PhysRevB.65.144508,fujimoto2007fermi,PhysRevLett.92.097001,yoshizawa2021atomic}, and superconducting states with a spatially modulated order parameter\cite{PhysRevLett.89.227002,PhysRevLett.94.137002,PhysRevB.76.014522,PhysRevB.92.014509}, etc. However, the intriguing Rashba effect on iron-based superconductors (IBSs) has not been observed so far. 

\begin{figure*}[htp]
       \begin{center}
       \includegraphics[width=6.5in]{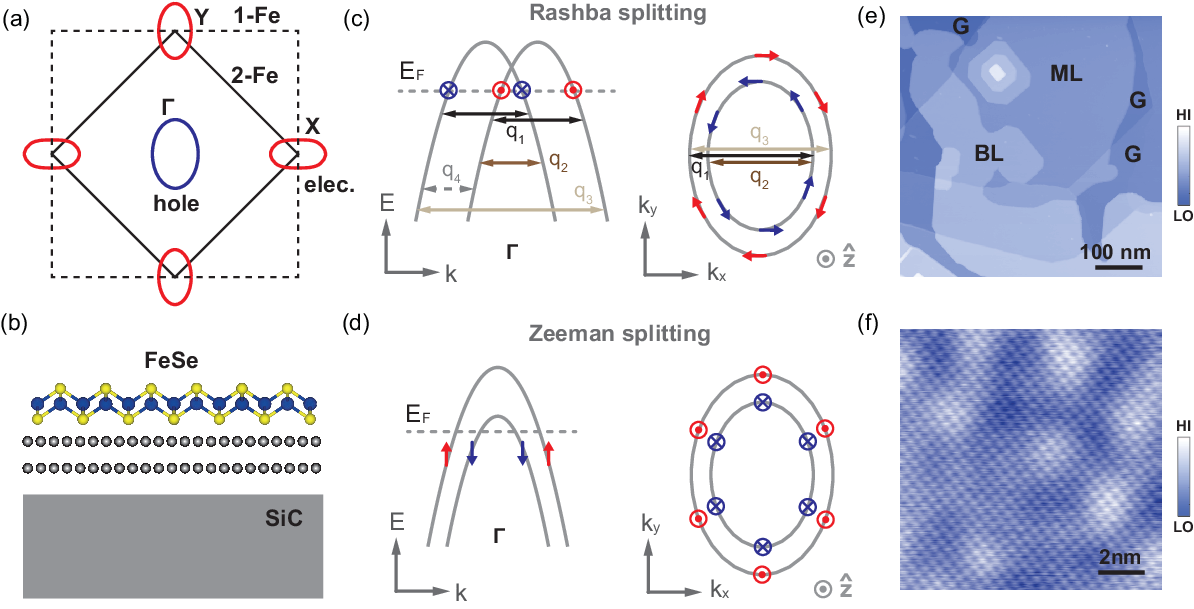}
       \end{center}
       \caption[]{
(a)  Schematic of the Brillouin zone and the Fermi surface. In the nematic phase, the electronic structure should be more properly viewed in the two-Fe Brillouin zone. 
(b) Side view of the FeSe  monolayer grown on a bilayer graphene (BLG)/6H-SiC(0001) substrate.
(c) Spin-split hole bands (left) and spin structures (right) due to Rashba-type SOC. The Fermi energy $\it E =  E_F$ is denoted. Solid bidirectional arrows represent intraband ($q_1$) and interband ($q_2$ and $q_3$) scattering vectors as observed in the experiment. The dashed arrow marks the scattering vectors that is not observed in the experiment. The in-plane spin orientations are indicated in blue and red arrows. 
(d) Spin-split hole bands (left) and spin structures (right) due to Zeeman-type SOC. The spins are oriented in the out-of-plane directions. 
(e) Typical topographic image of FeSe films grown on a BLG/SiC substrate, including FeSe monolayer, bilayer, and bare graphene on the surface. (490 nm $\times$ 490 nm, Setpoint: {\it V} = 3 V, {\it I} = 20 pA). 
(f) Atomically resolved STM topography of FeSe monolayer. (10 nm$ \times$ 10 nm, -30 mV, 0.1 nA). 
}
\label{fig:fig1}
\end{figure*}

Among IBSs, FeSe stands out due to its simple crystalline structure \cite{Paglione10} and the enhanced superconducting transition temperature $\it T{_c}$ in monolayer films \cite{Xue12}. FeSe undergoes a tetragonal-orthorhobic structure transition \cite{Wu08} at $\sim$90 K, stabilizing a nematic electronic state without long-range magnetic order\cite{PhysRevX.6.041045}. In the orthorhombic phase, the Fermi surface of bulk FeSe consists of ellipsoidal hole pockets around $\Gamma$=(0,0) point and electron pockets around X=($\pi$/$a_{Fe}$,0) and Y=(0,$\pi$/$a_{Fe}$) points [Fig. ~\ref{fig:fig1}(a)]. For FeSe monolayers grown on graphene/SiC substrate [Fig. ~\ref{fig:fig1}(b)], the band structure resembles the bulk but with an extremely low Fermi energy ($<$5 meV)\cite{huang2021superconducting}. Due to the broken inversion symmetry in the normal direction, the band degeneracy is lift, resulting in two Fermi surfaces with chiral spin polarization [Fig. ~\ref{fig:fig1}(c)]. In contrast, Zeeman-type SOC, caused by in-plane inversion symmetry breaking, leads to the out-of-plane spin polarization [Fig. ~\ref{fig:fig1}(d)]. These unique spin structures have remarkable implications for superconductivity under strong magnetic fields \cite{PhysRevLett.87.037004,PhysRevLett.111.057005,lu2015evidence,nam2016ultrathin}. In our case, measuring the meV-scale pockets and sub-meV band splittings is a challenge. Although laser-based angle-resolved photon-electron spectroscopy (ARPES) \cite{PhysRevX.8.031033} can achieve meV resolution, it only exhibits a limited momentum space around the $\Gamma$ point and achieving a resolution better than 1 meV is difficult. In this study, we offer spectroscopic evidence for band splitting in an FeSe monolayer utilizing ultra-low temperature scanning tunneling microscopy (STM) and spectroscopy (STS). The high resolution of STS enables us to discern $d$-orbital Landau quantization as well as the splittings of hole and electron bands.

\begin{figure*}[htp]
       \begin{center}
       \includegraphics[width=6.5in]{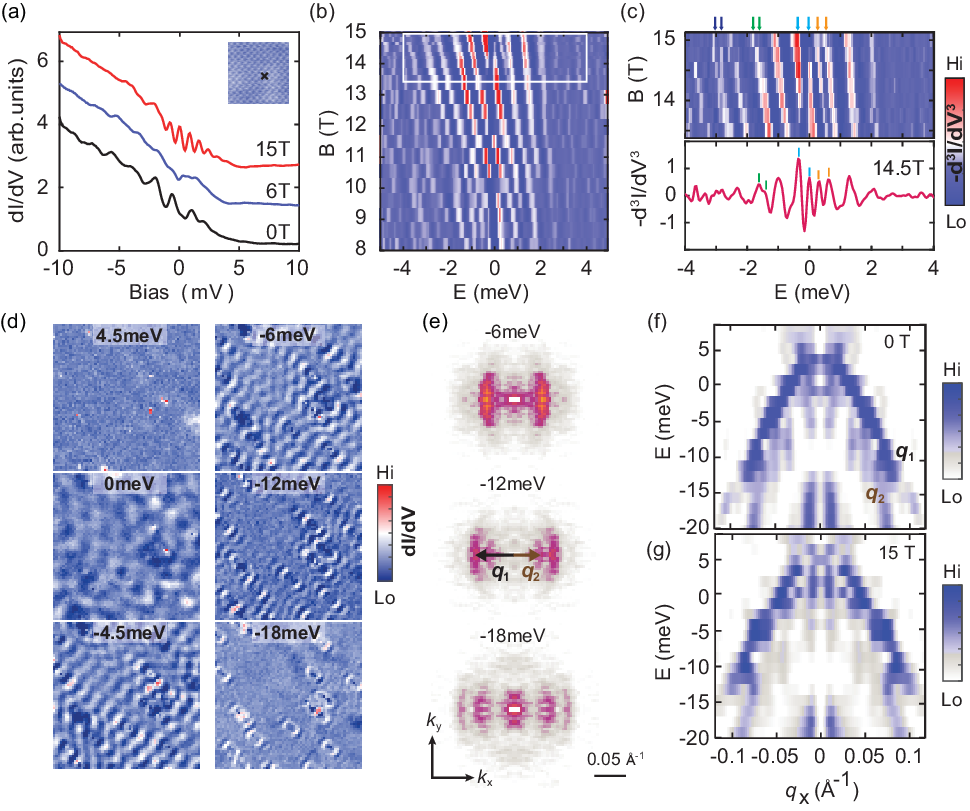}
       \end{center}
       \caption[]{
(a) dI/dV spectra at 0, 6 and 15 T ({\it V} = 10 mV, {\it I} = 100 pA, {\it V}$_{mod}$ = 0.1 mV). The curves are offset for clarity. Inset: 5 nm $\times$ 5 nm area in domain 1, with data taken at the location marked by the black cross. (b) Landau fan diagram showing hole-like LLs. The diagram is constructed from -d$^3I$/d$V^3$ (B,E) spectra data taken in domain 1 ({\it V} = 10 mV, {\it I} = 100 pA, {\it V}$_{mod}$ = 0.1 mV). (c) Zoom-in on the Landau fan diagram marked by a white box in (b), with additional splittings of LLs indicated by colored arrows. (d) dI/dV maps at different energies (60 nm × 60 nm, {\it V} = -20 mV, {\it I} = 100 pA, {\it V}$_{mod}$ = 0.2 mV), acquired over the same area shown in the inset of (a) but covering a larger area. (e) Fourier transform images of the dI/dV map of FeSe monolayer at -6, -12, and -18 meV (see more in Fig. S4). The data were symmetrized taking advantage of the mirror symmetry axes of the orthorhombic crystal unit cell. The size of two vectors $q$ is labeled $q_1$ and $q_2$, as indicated in Fig. 1(c). (f,g) QPI dispersion obtained by cutting lines along $k_x$. The dispersions are fitted with Eq.\ref{en-spin-orb1} to obtain the splitting parameter $\alpha\approx$ $46.5\pm5.3$ meV \AA{} at 0 T (f) and $\alpha\approx$ $49.0\pm6.1$ meV \AA{} at 15 T (g). Further details are provided in the supplementary materials. 
}
\label{fig2}
\end{figure*}

A series of FeSe monolayer films with different superconducting gap sizes \cite{huang2021superconducting} were fabricated on a SiC substrate. The typical lateral size of an FeSe island is about 200 nm (Fig. 1(e)).  The atomically resolved STM image in Fig. ~\ref{fig:fig1}(f) reveals the Se atoms in the top layer. The lattice constant of 3.75 \AA{}  agrees well with the bulk value. The STS at B=0 T in Fig. 2(a) (see also Fig. S1) shows the LDOS of the FeSe monolayer, indicating quenched superconductivity due to the low carrier density\cite{huang2021superconducting}. The peak feature around the Fermi level is attributed to the lateral quantum confinement (Supplemetal Material Section 4). These peaks are suppressed by elevated temperature [Fig. S1(a)] and magnetic fields [Fig. S1(b)]. The perpendicular magnetic field forces the motion of electrons into the cyclotron orbits. When the size of the orbit is comparable to the potential well width, the cyclotron motion surpasses the quantum well states and the quantum confinement effect is attenuated. As a result, the quantized peaks start to be suppressed at 6 T [blue curve in Fig. \ref{fig2}(a) and Fig. S1]. Subsequently, Landau quantization becomes visible in the energy spectra [red curve in Fig. \ref{fig2}(a)].

The high quality of the thin films allows us to investigate the band structure using Landau quantization and quasiparticle interference (QPI). Figure 2(b) illustrates the Landau fan diagram assembled from the spectra measured from 8 to 15 T at a fixed location in domain 1 [inset of Fig. 2(a)]. The linear dependence of LLs on the magnetic field, with negative slope, suggests a parabolic hole-like band dispersion. The quantized states span $\pm$5 meV and the LLs are equally spaced below 13.5 T. The spacing between two adjacent LL peaks is roughly half of the calculated Zeeman energy (Fig. S2). For instance, the calculated Zeeman energy at 13 T is 1.5 meV (g = 2\cite{PhysRevB.107.104517}), while the experimental LL spacing is 0.7 meV. There are two possible explanations for this discrepancy: (i) the Rashba field forces the spins in the in-plane direction and the LL spacing is determined by the orbital cyclotron energy $\hbar\omega_c$ rather than Zeeman energy, where $\omega_c$ is the cyclotron frequency; (ii) each LL$_n$ (for n$\geq$2) with index (n,s)(where n is the orbital index and s is $\pm$1/2 for spin up and down) coincides with LL$_{n-2}$ [Fig. S2]. The latter explanation can be dismissed by closely examining the diagram from 13.5 to 15 T [Fig. 2(c)]. In this range, certain LLs near $E_F$ are split into two branches, as indicated by colored arrows. The intense magnetic field competes with Rashba field and likely tilts the in-plane spins, meaning the spins are not entirely polarized by the external field due to Rashba SOC, resulting in the additional splitting observed in the LL spectra. However, such additional splitting is insufficient to allow for a detailed analysis of the Rashba parameters. Notably, the hole surface density can be determined to be $\sim$6$\times$10$^{12}$ cm$^{-2}$ from LLs (Supplemetal Material Section 1). This low density in the high-field regime may indicate strong many-body effects, similar to those observed in GaAs\cite{PhysRevB.45.8829,2007High}, graphene\cite{2010High}, bismuth\cite{Feldman2016Observation} and monolayer MoS$_2$\cite{PhysRevLett.121.247701}.

For the hole band, the splitting is clearly manifested through the analysis of electronic band dispersion. The differential tunneling conductance images [Fig. ~\ref{fig2}(d) and Fig. S3] in domain 1 of the FeSe monolayer at different energies reveal distinct standing wave patterns around Fe vacancy defects. These standing wave patterns exhibit strong in-plane anisotropy, which is attributed to the orbital selective coherence \cite{Davis18}. The dI/dV map at -4.5 meV shows unidrectional, rather than circular, LDOS oscillations resulting from quasiparticle interference associated with {\it d}$_{yz}$ orbital. This indicates that the quasiparticle states with {\it d}$_{yz}$ character are significantly more coherent than those with {\it d}$_{xz}$ character\cite{Davis18}. Figure ~\ref{fig2}(e) shows the Fourier transform of the QPI image at -6, -12, and -18 meV. Within the Fourier transform pattern, two intrapocket scattering wavevectors labeled q$_1$ and q$_2$ can be distinguished. Line cuts along $q_x$ at each energy are shown in Fig. ~\ref{fig2} (f), revealing two distinct parabolic dispersions. The length increases with decreasing energy, indicating a hole-band dispersion. Fitting the dispersion yields an effective mass of $m_x = 1.6 m_0$ along $x$-axis, which is smaller compared to the $m^* = 2.3 m_0$ derived from LL dispersion (Fig. S2(e)). This discrepancy arises from the anisotropic ellipse-like hole pocket, with different effective mass $m_x$ and $m_y$ along $x$- and $y$-axis. From $m^* = \sqrt{m_xm_y}$, we calculate $m_y = 3.26 m_0$, giving an ellipse ratio of 1.43, which is smaller than that of the bulk\cite{PhysRevX.8.031033}. Similar QPI-derived band splitting has been observed in bulk FeSe\cite{kasahara2014field} and FeSe$_{1-x}$S$_{x}$ \cite{Hanaguri2018Two} due to the different states of the hole bands $k_z$. 
However, our monolayer films lack the dispersion in the $z$-direction. According to the ARPES results, in the nematic phase, the energy splitting between the bands {\it d}$_{xz}$ and {\it d}$_{yz}$ around the point $\Gamma$ is up to 50 meV in bulk\cite{PhysRevB.91.155106,PhysRevB.90.121111,PhysRevLett.113.237001} and is of the same order of magnitude in single layers on SrTiO$_3$ substrates \cite{liu2021high,yi2015observation,PhysRevB.94.115153}. Therefore, we attribute these split bands to the Rashba spin-orbital interaction. Another indication of Rashba band splitting comes from the QPI intensity. In particular, the QPI intensity at $q_1$ is stronger than at $q_2$. This QPI intensity is proportional to the intensity of the spectral function in the initial states $k_1$ and the final states $k_2$, which can be substantially reduced if the spins at $k_1$ and $k_2$ are reversed, especially for QPI assisted by non-magnetic defects. The lower intensity for $q_2$ is consistent with Rashba SOC [Fig. 1(c)] rather than Zeeman SOC [Fig. 1(d)]. In our case, Fe vacancies are magnetic and can serve as the scattering centers for spin reversal. Such similar QPI branch associated with spin-flip scattering spin-flip scattering induced by magnetism has also been observed in topological insulators\cite{jack2020observation,bilayers2014one}.

To further rule out the Zeeman-type SOC, which arises from the breaking of in-plane inversion symmetry in 2D systems, a magnetic field perpendicular to the surface was applied. Zeeman-type SOC forces the spins to polarize in the out-of-plane direction, enabling interaction with the perpendicular magnetic field. If Zeeman-type SOC were present, a field of 15 T would induce an expected additional Zeeman splitting of 1.74 meV. Figure ~\ref{fig2}(g) shows the dispersion under 15 T acquired at the same location as that in Fig. ~\ref{fig2}(f). The fits yield a band splitting of 4.56 meV at 0 T and 4.77 meV at 15 T (Fig. S6). Therefore, the additional band splitting of 0.21 meV at 15 T is much lower than the expected 1.74 meV. This discrepancy suggests that Zeeman-type SOC is unlikely to be the dominant cause of band splitting at 0 T. In contrast, in the case of Rashba SOC, the Rashba field cannot be overcome to detect the significant Zeeman splitting (at 15 T). When fitting the dispersion in the case of Rashba SOC, Eq.\ref{en-spin-orb1}:
\begin{equation}
E(k)= \epsilon(k) \pm \alpha k   
\label{en-spin-orb1}
\end{equation}
the Rashba parameter is  $\alpha\approx$ $49.0\pm6.1$ meV \AA{} at 15 T, close to the value at zero field  $\alpha\approx$ $46.5\pm5.3$ meV \AA{} (Fig. S6). The magnitude of Rashba spin-orbit coupling varies in domains and can reach up to  $100$ meV \AA{} (Fig. S7). This variation may due to the strain in substrate or other factors during growth. 

\begin{figure*}[htp]
       \begin{center}
       \includegraphics[width=6.0in]{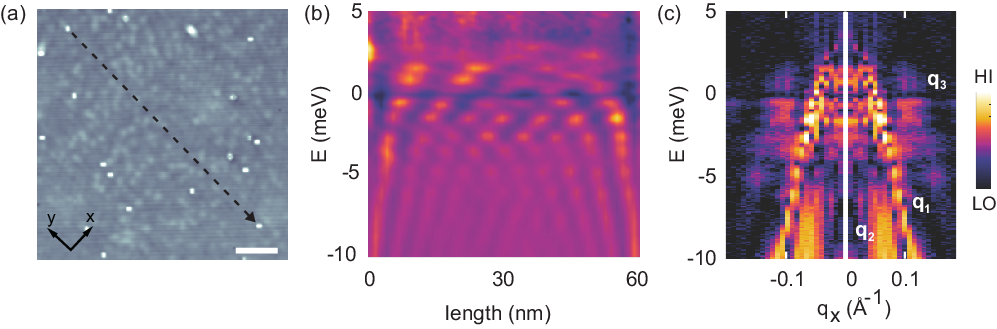}
       \end{center}
       \caption{
(a) Topographic image of domain 1 in FeSe monolayer, showing bright protrusions corresponding to Fe vacancies (60 nm $\times$ 60 nm, 100 mV, 100 pA, scale bar: 10 nm). (b) Spatially resolved dI/dV spectra measured along the dashed arrow in (a), which spans 60 nm. The arrow between two Fe vacancies is oriented parallel to y axis. All spectra have been subtracted from a linear background. ({\it V} = -10 mV, {\it I} = 0.1 nA, {\it V}$_{mod}$ = 0.1 mV). (c) Fourier transform of the spatially resolved dI/dV spectra in (b), with three scattering vectors $q_1$, $q_2$, $q_3$ labeled.
}
\label{fig:fig3_new}
\end{figure*}

In order to resolve more possible scattering wavevectors, we performed the spatially resolved dI/dV spectra along the dashed arrow between two Fe vacancies in Fig. \ref{fig:fig3_new}(a). This direction was chosen to be parallel to the y axis of the Se-Se lattice due to orbital selectivity. The line spectra show quantized peaks and standing wave pattern [Fig. \ref{fig:fig3_new}(b)], reminiscent of a particle-in-a-box, which can be modeled by using a Fabry-Perot resonator (see Supplemetal Material Section 4). Notably, a third scattering wavevector $q_3$ is identified from the Fourier transform of the line spectra [Fig. \ref{fig:fig3_new}(c) and Fig. S5]. All three scattering wavevectors show hole-like dispersion, corresponding to the $q$-vectors labeled in Fig. 1(c), which suggests the presence of the Rashba-SOC split Fermi surface.

\begin{figure*}[htp]
       \begin{center}
       \includegraphics[width=6.5in]{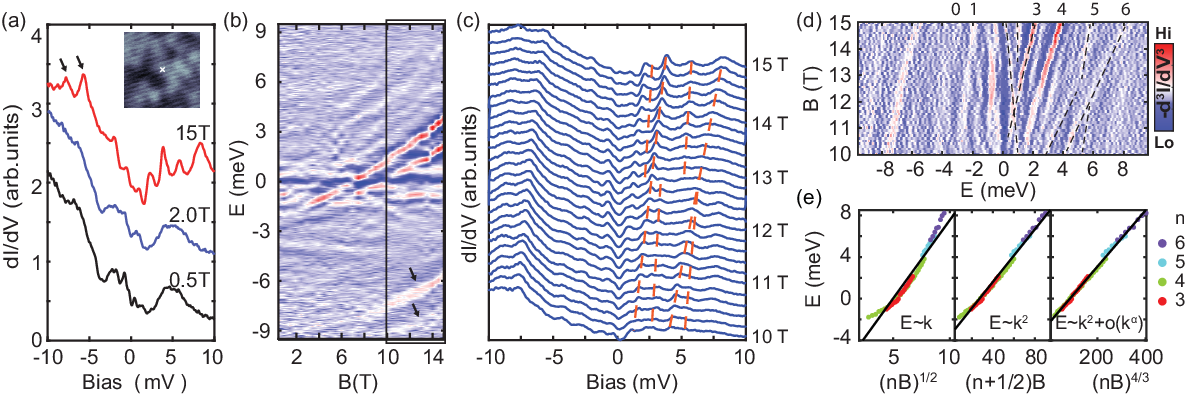}
       \end{center}
       \caption[]{
(a) dI/dV spectra at 0.5, 2 and 15 T ({\it V} = 10 mV, {\it I} = 100 pA, {\it V}$_{mod}$ = 0.1 mV). The curves are offset for clarity. Inset: Topographic image of a 10 nm $\times$ 10 nm area in sample $\#$3 with data taken at the location indicated by the white cross.   
(b) Landau fan diagram showing electron-like LLs. The diagram is constructed from -d$^3I$/d$V^3$ (B,E) spectra taken with 0.5 T increments from 0.5 to 15 T ({\it V} = 10 mV, {\it I} = 100 pA, {\it V}$_{mod}$ = 0.1 mV).     
(c) A series of dI/dV (B,E) spectra taken at the same location as in (a), ranging from 10 to 15 T in 0.2 T increments. The electron-like LL indexes are labeled. The orange dashed lines label the ``butterfly'' pattern.
(d) Zoom-in on the Landau fan diagram indicated by a black box in (b). The data represent a negative second derivative of the data in (c). The labeled index $n$ are labeled according to the model fits in Fig. 5. (e) LL peak energies extracted from (b) are plotted as functions of $(nB)^{1/2}$, $(n+1/2)B$ and $(nB)^{4/3}$. A linear fit is applied to analyze the dependence. For n = 3, the LL peak energies are acquired from 6.5 to 15 T; for n = 4, rom 3 to 15 T; and for n = 5 and 6, from 11.5 to 15 T.   
}
\label{fig3}
\end{figure*}

\begin{figure*}[htp]
       \begin{center}
       \includegraphics[width=6.25in]{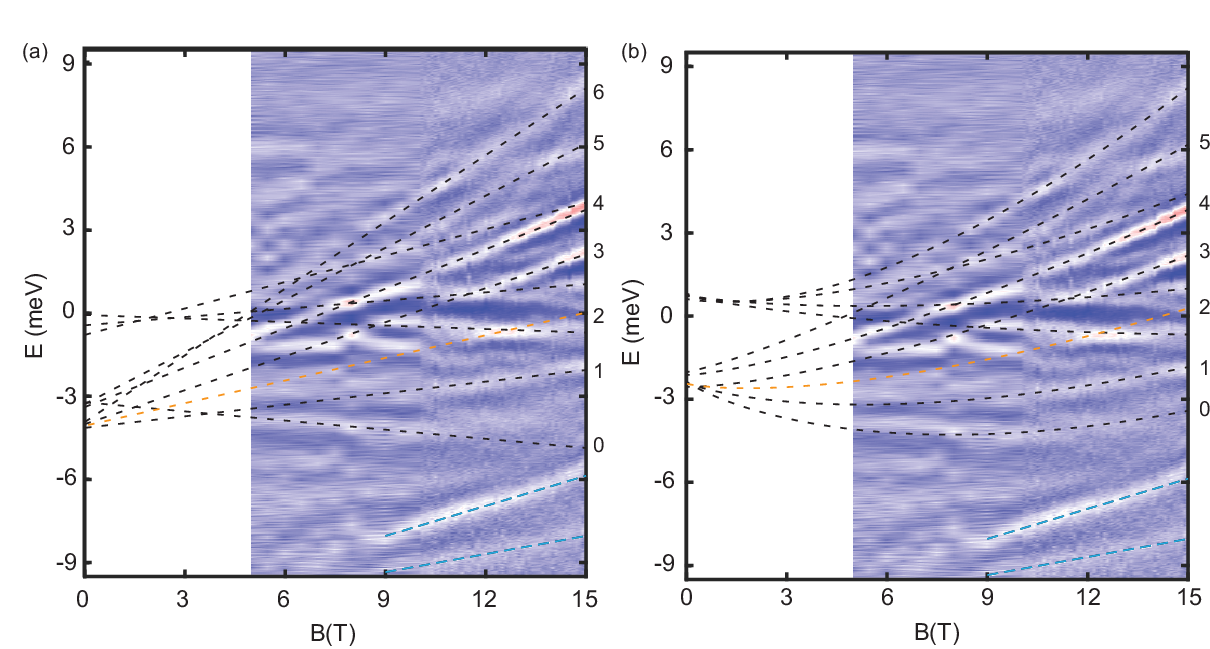}
       \end{center}
       \caption{(a) Fitting results for each LL using a linear function. The black dashed lines are the fits and extrapolated to zero field. The indices LL$_0$, LL$_1$ are labeled directly, while LL$_n$ (n = 3, 4, 5, 6) are assigned based on the fitted slope relative to that of LL$_1$. For instance, the slope of LL$_3$ divided by that of LL$_1$ is 2.95, leading to its assignment as LL$_3$. The other unindexed branches correspond to LLs of the upper band. (b) Fitting results for each LL using the formula $E = aB^{\frac{4}{3}} + bB + c$. One set of LLs converges at 0.75 $\pm$ 0.09 meV at 0 T, while the other set converges at -2.29 $\pm$ 0.21 meV, corresponding to the bottoms of two electron bands. The LL$_2$, indicated by orange dashed line, is calculated based on the fitted parameters from other LLs. Blue dashed lines indicate the LLs from the deeper electron band. 
       }		
\label{fig5}
\end{figure*}

For extremely shallow electron bands\cite{huang2021superconducting}, we resort to the band structure information from Landau-level spectroscopy. To resolve the LLs of the electron band, an extremely sharp STM tip is needed to access the electronic states at high k near the X/Y point\cite{PhysRevB.107.104517}. Figure \ref{fig3}(a) shows dI/dV measurements obtained using a sharper tip at 0.5, 2 and 15 T compared to the hole band measurement, which show similar behavior: the quantized peaks are first quenched before the formation of LLs as the magnetic field increases. From the Landau fan diagram [Fig. \ref{fig3}(b) and Fig.\ref{fig5}], two distinct sets of LLs are clearly observed, both displaying electron-like behavior, as inferred from their positive slopes. One set of LLs (marked by black arrows in Figs. \ref{fig3}(a) and (b)) at energies lower than -6 meV converges to -11.3 meV at zero field, indicating the bottom of a deeper electron band. The other set of LLs near the Fermi level consists of at least 8 peaks and exhibits non-linear dependence on the magnetic field. Zooming into a magnetic field ranging from 10 T to 15 T [Fig. \ref{fig3}(c-d)], some LLs cross each other and form a butterfly-like pattern. This feature is consistently observed at different locations. To comprehend these LLs, we extracted the peaks positions, $E_n$ at each magnetic field and examined the relationship between $E_n$ and $nB$ (or $(n+1/2)B$). We observe that the LL's dependence on the magnetic field preferably follows a $\sim (nB)^{4/3}$ relationship, as illustrated in Fig. \ref{fig3}(e). This nonlinearity observed in the Landau spectra suggests the presence of high-order effects in the electron band, such as a $k^4$-term. Given the orthorhombic symmetry of FeSe and the presence of a $k^4$-term, a substantial transformation in the shape of the electron band would be expected.

To gain further insight into the Landau-level structure of the shallow electron bands, we develop a model based on the Luttinger method. This approach employs basis functions derived from irreducible representations of the symmetry group at a $k$-point in the Brillouin zone\cite{bir1974symmetry}. It is particularly effective in describing the electron bands near the M or X/Y points of the Brillouin zone \cite{dresselhaus2007group,KUSMARTSEV19891}, ensuring a predominant electron character with a positive effective mass and corresponding to the point group of these high-symmetry points. This approach allows for a systematic expansion of the energy bands in terms of momentum components, constrained by the symmetry of the system; for example, the crystal and the k-point symmetry group as $D_{4h}$. To achieve this, it is necessary to also incorporate higher-order symmetry invariants, such as in the simplest form $(k_x^2+k_y^2)^2=k^4$ and $k_x^4+k_y^4$, resulting in the effective Hamiltonian:
\begin{equation}
H= I M_{0} + \tau_x   M_{1} +  \tau_z  M_{2}
\label{Ham-sym-inv}
\end{equation}
here \textit{I} is unit and $ \tau_x $  and $ \tau_ z$  are Pauli matrices, and
\begin{equation}
 M_0= b  \left( k_{x}^{2}\mathrm{+}k_{y}^{2} \right) +b_2  \left( k_{x}^{2}\mathrm{+}k_{y}^{2} \right)^2 
 + \beta (k_x^4+k_y^4) + ~ A_0\varepsilon _{zz}+B_0~ \left(  \varepsilon _{xx}+ \varepsilon _{yy} \right) 
\label{M0}
\end{equation} 
\begin{equation}
M_{1} = C k_{x}k_{y} +  C_0 \varepsilon _{xy}  
\end{equation}
\begin{equation}
M_{2} =D( k_{x}^{2}-k_{y}^{2})+ D_0 (  \varepsilon _{xx}- \varepsilon _{yy}) 
\end{equation}

Where \textit{k\textsubscript{x}, and k\textsubscript{y}},\  are values of the wavevector counted from the chosen M or X point of the BZ. The parameters \textit{$B, C, D$} and \textit{$A_0,~B_0,~C_0,~D_0$} are arbitrary constants whose values should be determined experimentally. Here, \(  \varepsilon _{xx},~ \varepsilon _{yy},~ \varepsilon _{zz}, ~\varepsilon _{xy} \) \textsubscript{ } are lattice strain tensor components, which are relevant for FeSe transition to the nematic state with orthorhombic symmetry. \par


Here we present a simple analytic model for Landau and cyclotron $ B^{4/3}$ quantization where we explicitly include nonlinear terms like $k_x^4+ k_y^4$.
In the presence of a uniform magnetic field \( \mathbf{B} \), the Hamiltonian describing the motion of an electron with charge \( e \) and mass \( m \) is:
\[ H = \frac{1}{2m} \left(\mathbf{p} + e \mathbf{A} \right)^2 + \beta ( (p_x+e A_x)^4+(p_y+e A_x)^4) \]
where \( \mathbf{p} \) is the momentum of the electron, \( \mathbf{A} \) is the vector potential related to the magnetic field via \( \mathbf{B} = \nabla \times \mathbf{A} \). Choosing the \textit{Landau gauge} where \( \mathbf{A} = (-By, 0, 0) \), we can split the Hamiltonin considering x and y dependent components separately.
 The energy levels resulting from this quantization can be quasi-classically obtained by considering the total energy associated with the y-component Hamiltonian:
 \[ E = \frac{1}{2m} \left(p_x^2+ \frac{\hbar^2 n^2}{L^2} +\frac{e^2 L^2 B^2}{c^2} \right) + \beta  \left( \frac{\hbar^4 n^4}{L^4} +\frac{e^4 L^4 B^4}{c^4} +p_x^4+ 2 p_x^2\frac{e^2 L^2 B^2}{c^2}\right) \]
 The Landau quantization we obtain by minimizing the energy with respect to the  parameter $L$. For example,  when $\beta$ vanishes, the minimization yields the value $L^4=\frac{c^2 \hbar^2 n^2}{e^2 B^2}$. Substituting this value for $L$ gives a conventional Landau quantization: 
\(  E_n = \hbar \omega_c n \). Here, the vacuum energy $\hbar \omega_c/2$ has been  neglected, 
and \( n = 0, 1, 2, \dots \) is the \textit{Landau level index} (a non-negative integer),
\( \omega_c \) is the \textit{cyclotron frequency}, given by \(  \omega_c = {eB}/{mc} \),
\( \hbar \) is the reduced Planck constant,
\( e \) is the electron charge,
\( B \) is the magnetic field strength,
\( m \) is the electron's effective mass.
When the value $\beta$ is large, the minimization with respect to $L$ yields $L^2=(\frac{\hbar^2 n^2 c^4}{4m e^4 B^4})^{1/3}$, leading to a $B^{4/3}$ quantization:
\begin{equation}
E_n =c_1 (nB)^{4/3} (\frac{\beta e^4 \hbar^4 }{m^2 c^4})^{1/3}
\end{equation}
where $c_1$ is a constant about 1. The complete solution of this model is obtained using Cardano formula and provided in the supplementary materials. The LLs now are described by the conventional Linear Landau  quantization term and a new term $\sim (nB)^{4/3}$, as explained in this model and observed in our experiments. 

Figure \ref{fig5}(a) shows the fitting results using the standard linear function $E=aB+b$ for each LL$_n$. In contrast, the fits that include the additional non-linear term $\sim B^{4/3}$ are shown in Fig. \ref{fig5}(b). In both cases, the fits (black dashed lines) converge to two values at zero field, corresponding to the bottoms of two electron bands at X/Y points. However, for the linear fits, the values converging at 0 T exhibit larger error bars and the fits deviate from the experimental data in high field regime. Comparatively, the fits that include both $B$ and $B^{4/3}$ term (Fig. \ref{fig5}(b)) align well with the experimental data, suggesting the presence of $k^4$ term in the electron band. Although LL$_2$ (orange dashed lines) is less distinct, which may be due to complex LL crossings and many-body interactions near the Fermi level, it has been calculated based on the fitted parameters from other LLs. Beyond this simple analytic model, we propose a more comprehensive framework based on the Luttinger method (Supplemental Material Section 7), which reveals that the electron band around the X/Y point in FeSe monolayers resembles a distinctive, flattened double potential well [Fig. S8(c)].


In conclusion, we have investigated the Rashba SOC effect on the band structure of FeSe monolayer through high-resolution LL spectroscopy and QPI measurements. The interplay between Rashba and Zeeman interactions further splits the LLs and potentially results in a mixture of spin-singlet and triplet Cooper pairing or more intriguing quantum states, which needs more investigations. 

For the electron band at the X/Y point, we observe a non-linear dependence of each LL on the magnetic field, manifesting itself as the behavior $B^{4/3}$ instead of the conventional linear dependence on B. The deviation indicates the dominance of the $k^4$-term in the electron band. Our calculations, based on the symmetry analysis with the $k$-point model, offer a phenomenological approach to incorporate lattice-strain effects and enable manipulation of the topological properties of this material. 

Additionally, our results highlight how LLs can serve as a probe of the topological nature of electronic bands\cite{PhysRevLett.126.056401}, opening new avenues for observing and exploring topological states in systems with extremely low carrier concentrations. Our findings indicate that external parameters such as strain or pressure, can significantly influence the topological properties, providing opportunities for material design tailored for technological applications.

\section{Materials and Methods}

The experiments were carried out on a low temperature ultra-high vacuum  (UHV, 1$\times$10$^{-10}$ torr) STM equipped with molecular beam epitaxy (MBE). The magnetic field is applied perpendicular to the sample surface. The STM head can reach a base temperature of 60 mK with an effective electronic temperature of 260 mK of the samples \cite{chen19}.  FeSe monolayer films were grown on the n-type 6H-SiC(0001) substrate (nitrogen-doped, resistivity 0.02-0.2 $\Omega\cdot$cm) held at 400$^\circ$C during growth. To reduce the coupling between FeSe and the substrate to the van der Waals type \cite{PhysRevB.84.020503}, the surface of SiC was graphitized to form bilayer graphene [Fig. ~\ref{fig:fig1}(b)] in advance by thermal desorption of Si from the topmost layers. FeSe growth was carried out under Se-rich conditions at a rate of about two monolayers per hour and was monitored by {\it in situ} reflection high-energy electron diffraction. Before STM imaging, the polycrystalline Pt-Ir alloy tip was modified and calibrated on a clean Ag(111) surface. The STS measures the differential conductance d$I$/d$V$, which is proportional to the local density of states (LDOS). The d$I$/d$V$ spectra were acquired by the standard lock-in technique with a modulation frequency of $f=887$ Hz. 

\section{Supporting Information} 

The Supporting Information is available free of charge.

\begin{itemize}
  \item Evolution of dI/dV spectra as a function of temperature and magnetic field
  \item dI/dV maps
   \item Fourier transform of dI/dV maps
  \item Particle-in-a-box
    \item Fitting wavevector
   \item Variation of Rashba SOC
  \item Theoretical model for Rashba effect
\end{itemize}

\section{Acknowledgement}

This work was supported by the National Natural Science Foundation of China (Grant No. 12074211, No. 12141403, No. 52388201,  No. 12161141009 and No. 11934001), and the Ministry of Science and Technology of China (Grants No.2023ZD0300500). F.K. acknowledges support from the Chinese Academy of Sciences President's International Fellowship Initiative (Grant No. 2024VMA0005); Khalifa University grants: FSU-2021-030/8474000371 and RIG-2023-028; RIG-2024-046 and RIG-2024-053; Environmental Sustainability Awards: 2023,2024, the EU H2020 RISE project TERASSE (H2020-823878).

\bibliography{main}

\end{document}